\documentclass[12pt]{JHEP3}
\usepackage{graphics}
\usepackage{cite}

\def\to{\rightarrow}
\newcommand{\ket}[1]{|#1\rangle}
\newcommand{\bra}[1]{\langle#1|}

\def\ie{{\it i.e.\ }}
\def\leftrightarrowfill{$\m@th \mathord\leftarrow \mkern-6mu 
 \cleaders\hbox{$\mkern-2mu \mathord- \mkern-2mu$}\hfill
 \mkern-6mu \mathord\rightarrow$}
\def\overleftrightarrow#1{\vbox{\ialign{##\crcr
     \leftrightarrowfill\crcr\noalign{\kern-1pt\nointerlineskip}
     $\hfil\displaystyle{#1}\hfil$\crcr}}}
\newcommand{\beq}{\begin{equation}}
\newcommand{\eeq}{\end{equation}}
\newcommand{\bea}{\begin{eqnarray}}
\newcommand{\eea}{\end{eqnarray}}

\newcommand{\E}{{\cal E}}
\renewcommand{\d}{\delta}
\newcommand{\D}{\Delta}

\renewcommand{\a}{\alpha}

\newcommand{\m}{\mu}

\newcommand{\s}{\sigma}
\newcommand{\vx}{{\bf x}}

\newcommand{\oh}{\frac{1}{2}}
\newcommand{\dg}{\dagger}
\newcommand{\non}{\nonumber}

\newcommand{\rf}[1]{(\ref{#1})}
\newcommand{\ra}{\rightarrow}
\newcommand{\pa}{\partial}
\title{
{\small\rm\begin{flushright}
LBNL-49311\\
UFIFT-HEP--01-26
\end{flushright}
}
Gluon Chain Model of the Confining Force}

\author{Jeff Greensite \\ Physics and Astronomy Department, San Francisco
State University \\
San Francisco, CA 94117. 
E-mail: \email{greensit@quark.sfsu.edu} \\
{\rm and} \\
Theory Group, Lawrence Berkeley National Laboratory \\
Berkeley, CA 94720}

\author{Charles B. Thorn\footnote{Visiting Miller
Research Professor, on sabbatical leave from 
Department of Physics, University of Florida, 
Gainesville FL 32611. } \\ 
Department of Physics, University of California \\
Berkeley, CA 94720. Email: \email{thorn@phys.ufl.edu}\\
{\rm and} \\
Theory Group, Lawrence Berkeley National Laboratory \\
Berkeley, CA 94720}

\abstract{We develop a picture of the QCD string as a chain
of constituent gluons, bound by attractive nearest-neighbor forces
which may be treated perturbatively. This picture accounts
for both Casimir scaling at large $N_c$, 
and the asymptotic center dependence of
the static quark potential.  We discuss the relevance, to the
gluon-chain picture, of recent three-loop results for the static quark
potential. A variational framework is presented 
for computing the minimal energy and wavefunction of
a long gluon chain, which enables us to derive both 
the logarithmic broadening of the QCD flux tube (``roughening''), and 
the existence of a L\"uscher $-c/R$ term in the potential.}


\preprint

\begin{document}

\section{Introduction}

Many years ago \cite{thornweeparton,thornrpa,Jeff,GH} we suggested
a picture of the formation and 
composition of the QCD string as a linear chain of gluons,
which are the perturbative excitations the theory.  
This picture is motivated by 't Hooft's large-$N_c$ expansion. 
In particular, a time-slice of a high-order planar
diagram for a Wilson loop (Fig.\ \ref{planar}) reveals a sequence of gluons, 
each of which interacts only with its nearest 
neighbors in the diagram by mainly attractive forces.  This immediately
leads to the idea that the QCD string is composed of a ``chain'' of constituent
gluons, each held in place by its attraction to its two nearest neighbors
in the chain. The challenge in such a model is to understand how the
attractive force between gluons, which is essentially due to one-gluon
exchange, can manage to hold these massless constituent 
particles in a bound state. 
In this article we address this problem, and
in particular we show how recent perturbative results
for the static quark potential, and the related 
force renormalization scheme \cite{NS}, bear on this issue. 
We propose an appealing way to interpret
perturbative QCD that leads to a self-consistent
extrapolation of the perturbative force of a
quark on an anti-quark to a linear confining force. 
We do not claim that
this in any way {\it proves} confinement, but rather that
it provides a model framework for thinking about 
the physics of confinement
which stays conceptually close to perturbation theory.
A computational scheme is presented, involving variational and perturbative
elements, which we hope to 
eventually apply to calculate the ground state (and the corresponding static
quark potential) of the gluon chain.
Even without a full calculation of this ground state, we are able to use
our variational framework to demonstrate both the logarithmic 
broadening of the QCD flux tube, and the existence of a L\"uscher $-c/R$ 
term at long distances.

   Section 2, below, reviews the motivation of the gluon-chain
model, with emphasis on how this model accounts for features of
the confining force which are problematic for other theories of quark
confinement.  In section 3 we discuss in more detail the physics
of gluon chain formation, and the binding of gluons in the chain.  
The running coupling is, of course, crucial to this dynamics, and at 
the coupling strengths relevant to chain formation there are  
important issues of renormalization scheme 
dependence that must be confronted. In section 4 we present a
variational framework for semi-perturbative 
calculation, and show how it applies to a string tension calculation.
In this section we demonstrate how the gluon-chain model accounts for
roughening, and the L\"uscher $-c/R$ term in the static quark potential.
The last section contains some concluding remarks.%\\

\FIGURE[h!]{
\centerline{\scalebox{0.55}{\includegraphics{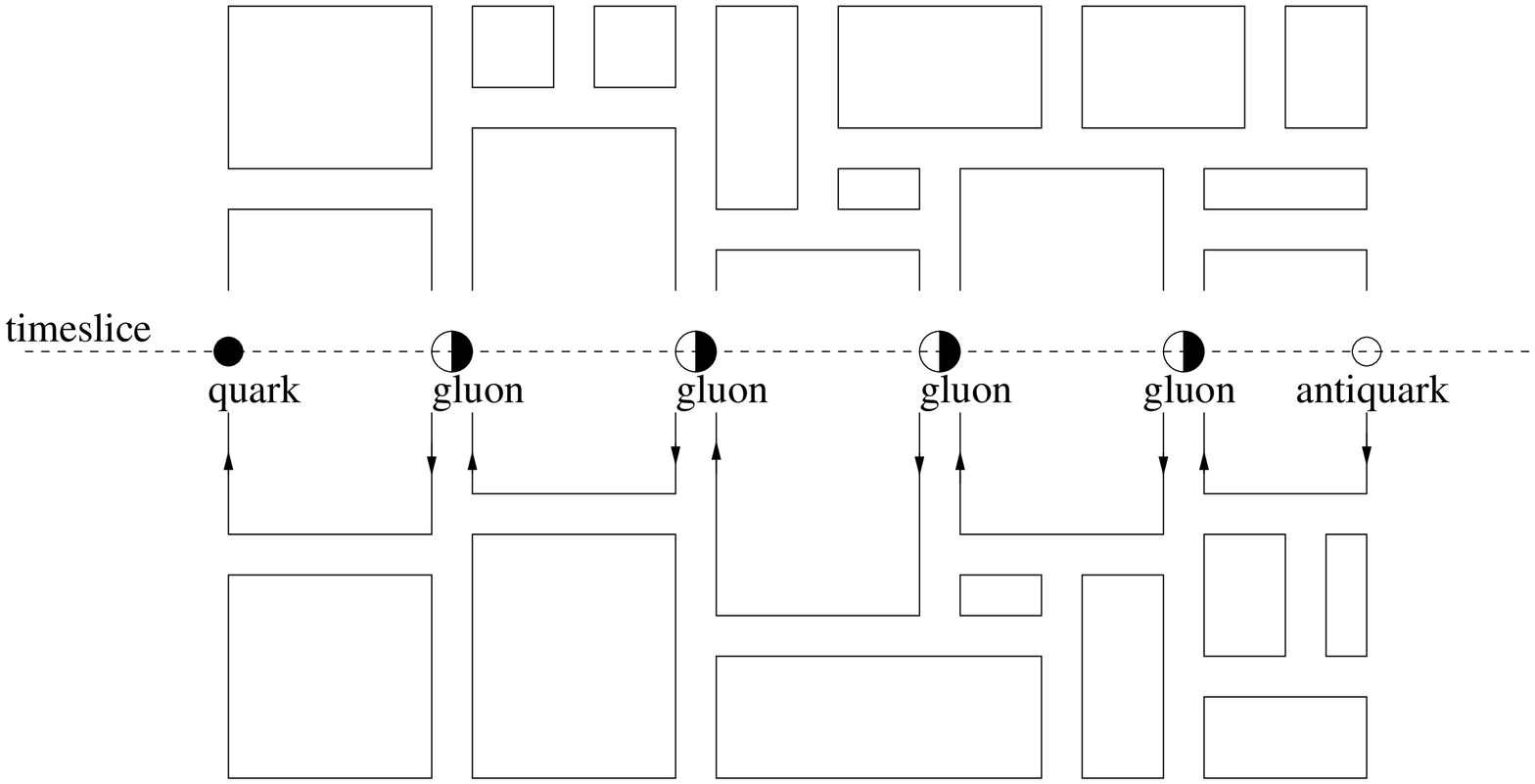}}}
\caption{The gluon chain as a time slice of a planar diagram (shown here
in double-line notation).
A solid hemisphere indicates a quark color index, open hemisphere
an antiquark color index.}
\label{planar}
}

\section{The Gluon-Chain Model}  

    Like any theory of quark confinement, the gluon-chain model aims at 
explaining the linearity of the 
static quark potential.  However, it is now widely 
recognized that in addition to the linearity feature, there
are at least three other properties of the confining potential which   
a satisfactory theory of confinement is obligated to explain:
\begin{itemize}
\item {\bf Casimir Scaling:~} Consider the potential between static
quarks in a representation $r$ of the gauge group.
From the onset of linearity in the potential,
to a finite (adjoint string-breaking) scale, 
the string tension $\s_r$ is proportional to the quadratic
Casimir $C_r$ of the representation, i.e.\
\beq
      \s_r = {C_r \over C_F} \s_F
\eeq
where the subscript $F$ denotes the fundamental representation.
For rectangular $L\times T$ Wilson loops with $L/T$ fixed, the range
of $L$ for which the Casimir
scaling law is valid increases logarithmically with $N_c$.
\item {\bf Center Dependence:~} Asymptotically,  the string tension
can depend only on the N-ality of the group representation $r$, i.e.\
on its transformation properties under the center subgroup of the
gauge group.
\item {\bf String Behavior:~} The diameter of the color-electric flux tube
between static sources is believed to grow 
logarithmically with the separation $L$ of
the sources (roughening), and there is a $-c/L$ contribution to the asymptotic
potential (the L\"uscher term) which is due to quantum 
fluctuations of the QCD string, 
rather than Coulomb attraction of the quarks. 
\end{itemize}

   Taken together, this is a challenging set of conditions. 
The abelian monopole theory \cite{mpole}, for example, has a very hard time 
accounting for Casimir scaling \cite{Lat96}, as well as for the center
dependence of certain operators \cite{j3}. 
Instanton \cite{Suganuma} and meron \cite{Negele} 
mechanisms are consistent
with Casimir scaling, but have difficulties with center dependence.
The center vortex theory \cite{vortex} is in perfect accord with
center dependence, and it is at least roughly compatible
with Casimir scaling, as shown in ref.\ \cite{Cas}.  On the other hand,
the vortex theory does not really explain the high degree of accuracy of
the Casimir scaling rule, which has been revealed in
numerical simulations \cite{Bali}.
Finally, the string-like behavior of the QCD flux tube 
seems to pose problems for 
any theory of confinement based essentially on one-gluon exchange 
(e.g.\ the proposal of ref.\ \cite{Zwanziger}), as well as 
for the proposal of stochastic confinement \cite{Simonov}.

   The gluon-chain model of QCD string formation,
on the other hand, meets the conditions listed above in a rather simple
and appealing way, as we will discuss below.  

\subsection{Linear Potential}

\FIGURE[h!]{
\centerline{\scalebox{0.35}{\includegraphics{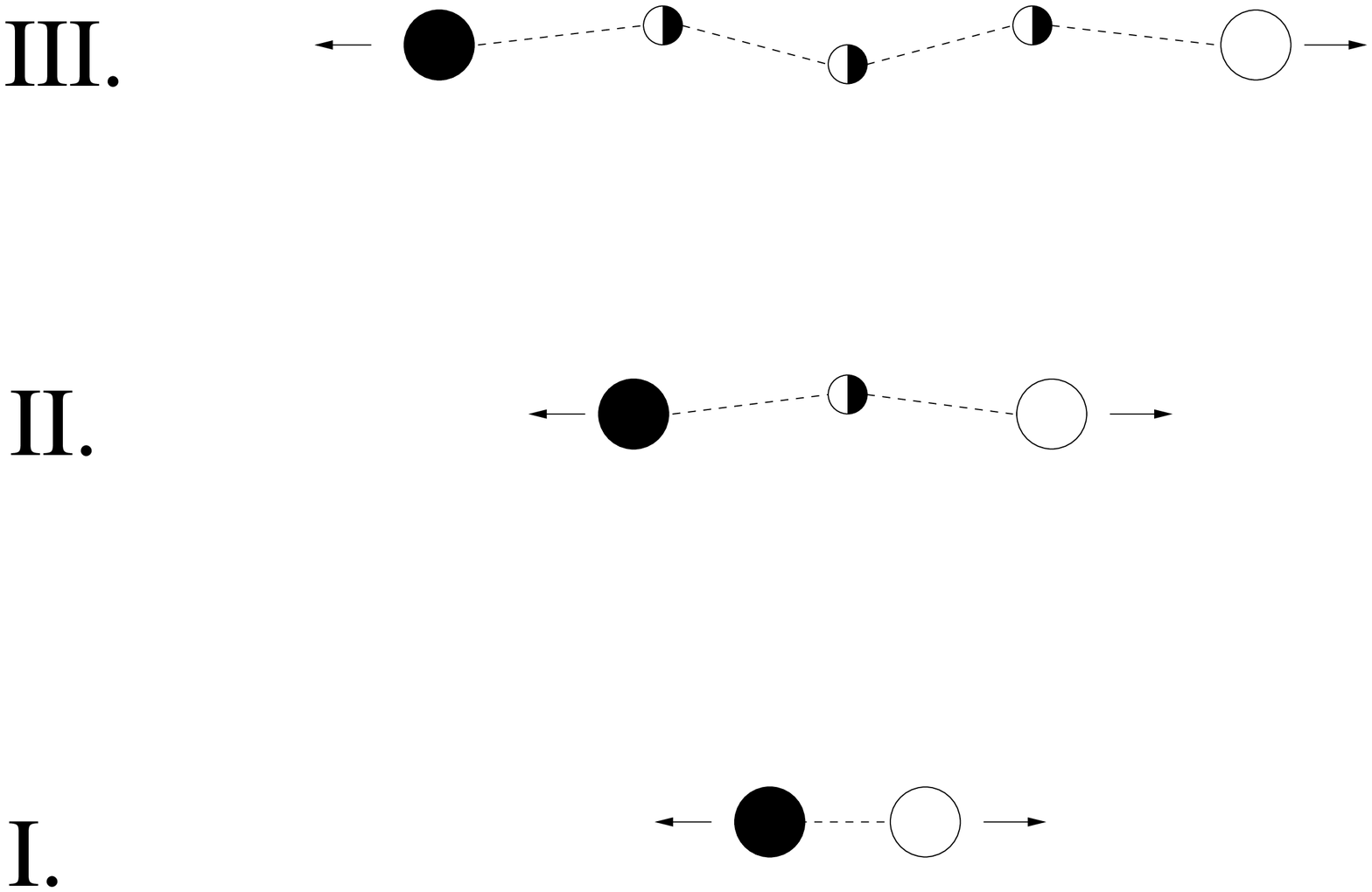}}}
\caption{As heavy quarks (large circles) separate, energy is 
minimized by keeping the average color charge separation below a certain
limit.  This is achieved by pulling out a sequence of gluons (small circles)
between the heavy quarks.  Again, solid and open shadings denote
quark and antiquark color indices, while dotted lines indicate contracted 
color indices between a quark and a gluon or between neighboring gluons.}
\label{single}
}

   As a heavy quark-antiquark pair move apart, 
and their color charge separation increases, we expect that at some
point the interaction 
energy increases rapidly due to the running coupling (cf.\ section 3).  
Eventually it becomes energetically favorable to reduce the
effective charge separation by inserting a gluon between the quarks.
In the $N_c\ra \infty$ limit, the quark and antiquark can only interact
with the intermediate gluon, but not directly with each other.
As the heavy quarks
continue to move apart, the process repeats, and we end up with a chain
of gluons, as shown in Fig.\ \ref{single},
in which the average distance
between color charges remains fixed, irrespective of the separation of
the heavy sources.  The energy of the system is approximately
$N E_{gluon}$, where $N$ is the number of gluons in the chain, and
$E_{gluon}$ is the kinetic and nearest-neighbor interaction energy
per gluon.  If the quarks are separated by a distance $L$, and the
number of gluons per unit quark separation ($N/L=1/R$) is fixed, then
\beq
      E_{chain} \approx N E_{gluon}
                = {E_{gluon} \over R} L
                = \s L
\eeq
where $\s=E_{gluon}/R$ is the string tension.  The linear growth
in the number of constituent gluons is the origin of the linear potential 
in the gluon-chain model.

   Alternatively, we may understand the linear potential in terms of
a constant force.  The force between neighboring constituent gluons
is dependent on the average separation $R=L/N$ of gluons along the
quark-antiquark axis.  If this separation remains fixed as $L$ increases,
then the average intergluon force remains fixed. The intergluon force, which
is the same everywhere in the chain, can be interpreted as a string
tension, which is constant irrespective of the quark separation. 

\subsection{Casimir Scaling}

   To leading order in $N_c$, a group character in representation
$r$ is given by a product of group characters in the fundamental 
representation 
\beq
      \chi_r[g] \propto \Bigl(\chi_F[g]\Bigr)^n 
                  \Bigl( \chi^*_F[g] \Bigr)^{\overline{n}}
           ~ + ~ \mbox{sub-leading terms} .
\eeq 
By factorization at large-$N_c$, a Wilson loop in representation $r$ has
a string tension
\beq
     \s_r = M_r \s_F
\eeq
at $N_c \ra \infty$, where $M_r = n + \overline{n}$.  In this limit,
the quadratic Casimir is $C_r = M_r N_c/2$.  Exact Casimir scaling is 
therefore a property of the planar limit.

   The gluon-chain model, which is motivated by large-$N_c$ considerations,
inherits this property.  A heavy source in representation $r$
is the terminus of $M_r$ separate gluon chains, one for each of the
$n$ quark and $\overline{n}$ antiquark charges in the direct
product forming the representation $r$.
Since the chains do not interact in the $N_c=\infty$ limit (the interaction
is a non-planar process, as can easily be verified by considering the
relevant Feynman diagrams), the total energy is simply the sum of the
energies of each of the chains.  In this way, Casimir scaling is obtained,
at least at large $N_c$.\footnote{At small $N_c$, the Casimir is not
simply proportional to $M_r$.  On the other hand, at small $N_c$, 
interactions between the chains cannot be neglected.  We cannot say, at
present, whether this model predicts any substantial deviation from Casimir 
scaling at small $N_c$.} 

\subsection{Center Dependence}

\FIGURE[h!]{
\centerline{\scalebox{0.35}{\includegraphics{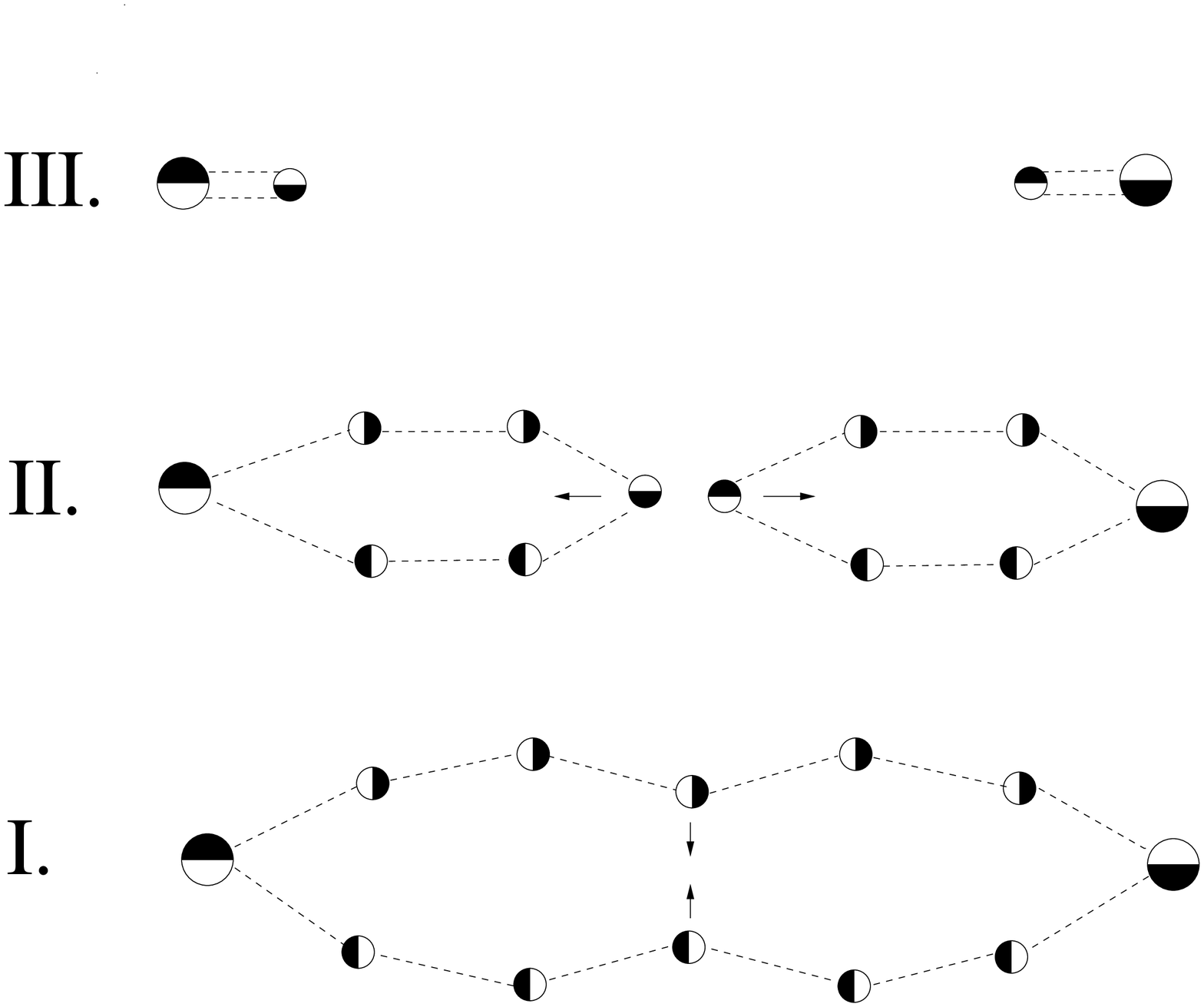}}}
\caption{Adjoint string-breaking in the gluon chain model.
Two gluons in separate chains (I) scatter by a contact interaction,
resulting in the re-arrangement of color indices indicated in II.
This corresponds to chains starting and ending on the same heavy source.
The chains then contract down to smaller ``gluelumps'' (III).}
\label{double}
}

   Asymptotically, the string tension of static sources in a 
higher color group representation $r$ depends only 
on the N-ality, rather than the quadratic Casimir, of the group representation.
Let us consider how this comes about in, e.g., the adjoint representation
(the analysis is easily generalized to other representations).  
Beginning with heavy quark sources
in the adjoint representation, there are two gluon chains, as shown in
Fig.\ \ref{double}.  Nearby gluons in each chain can scatter, e.g.\ by
a contact interaction, and rearrange the sequence of colors as shown
in the figure.  The result is that two gluon chains transform to two 
``gluelumps''; the adjoint string is broken (providing the sum of
gluelump masses is less than that of the double chain), and the resulting 
string tension is zero.  This is the correct prediction, since the N-ality
of the adjoint representation is also zero.  However, the scattering
process indicated is non-planar, and the transition rate from the two-chain
structure to the gluelump structure is $1/N_c^2$ suppressed.
In the large-$N_c$ limit, we therefore recover exact factorization
and Casimir scaling.

\subsection{String Behavior}

   It should be obvious, just from the figures, that a gluon-chain
is a discretized string \footnote{For a treatment of 
discretized bosonic strings in light-cone gauge, see ref.\ \cite{thornlcft}.}
of some kind, with the constituent gluons
playing the role of ``string-bits.''
Therefore it is reasonable to
expect that, due to quantum fluctuations of the chain configuration,
we should find the logarithmic broadening of the color-electric
flux tube with quark separation (roughening), as well as the L\"uscher
$-c/R$ term term in the static quark potential.  These effects are 
very non-trivial, however, and need to be
demonstrated in the context of the gluon-chain model.
We will postpone the analysis to section 4.  \\

\section{The Force Renormalization Scheme and Gluon Chain Formation}

   In the remarks above, we have passed lightly over a fundamental issue.
Gluons are massless particles.  The question is how nearest neighbor forces
between constituent gluons actually manage to bind such particles together
in a chain.  Clearly it is hopeless to find a binding mechanism at
very weak coupling $\a_s=g^2/4\pi$, 
since the kinetic energy of a gluon confined to
a region of size $R$ is of order $1/R$, whereas the interaction energy
(at tree level) is only of order $-\a_s/R$.  The fact that the
effective $\a_s$ is 
not really constant, but grows with distance scale $R$, is obviously 
of crucial importance.

  Superficially, a coupling which grows monotonically with distance seems 
exactly what is needed for quark confinement.  However, different
choices for what the coupling measures lead to drastically different
estimates of the quark-antiquark static potential.
For example, if we say that $\alpha_s(R)$ is a measure of the
potential energy between quark and anti-quark (the ``V-scheme''), 
we would then write the familiar expression for the potential
energy of a static quark and antiquark separated by distance $R$
in an SU($N_c$) gauge theory
\begin{eqnarray}
V_{q\bar q}(R)=-\left(1-{1\over N_c^2}\right){N_c\alpha^V_s(R)\over 2R}.
\label{Vscheme}
\end{eqnarray}
It should be understood that this formula is a {\it definition}
of the running coupling, and there is a complicated relation 
(beyond two loops) of $\alpha^V_s(R)$ in the V-scheme to %$\alpha_s(R)$ 
the running coupling in, 
e.g., the $\overline{\mbox{MS}}$ scheme \cite{NS,Bali1}.

    If $\a^V_s(R)$ grows monotonically with $R$, as suggested
by the first few terms in the renormalization group
Gell-Mann-Low function, the potential in eq.\ \rf{Vscheme}
is actually the \emph{opposite} of what is required for confinement.
Assuming that $\a^V_s(R)$ grows with $R$, perhaps even blowing
up at some finite $R_\infty$ (the Landau singularity), the potential in
eq.\ \rf{Vscheme} leads to a force which first weakens, then vanishes
at some point, and finally becomes ever more strongly repulsive.
The problem is that a monotonically increasing $\a^V_s(R)$ tends to
drive the potential away from the $V=0$ axis in the direction of 
negative $V$, whereas in fact
the static potential crosses the axis at some point and becomes
positive. Therefore, as defined in the V-scheme, $\a^V_s(R)$ derived from the
true static potential cannot possibly grow
monotonically.  Instead, at some scale, $\a^V_s$ must start to become smaller
and eventually change sign.  There seems to be little hope of relating
such behavior to the running coupling, at least up to three loops in
perturbation theory.   

%   It was realized many years ago by Thorn \cite{Charles1} and
%Mr.\ X \cite{mrx}\footnote{I misplaced the reference.  I'll have to
%look it up again.} 
As long advocated by one of us (C.B.T.), it is less problematic to
define the running coupling by the {\it force} (the derivative of
the potential) that the static quark exerts on the
static anti-quark \cite{mrx} 
\begin{eqnarray}
|F(R)|=\left(1-{1\over N_c^2}\right){N_c\alpha_s(R)\over 2R^2}
\equiv\left(1-{1\over N_c^2}\right){\pi\lambda(R)\over2R^2},
\end{eqnarray}
where we have defined the 't Hooft coupling held fixed in the $N_c\to\infty$
limit by $\lambda=N_c\alpha_s/\pi$.  In the rest of this
paper $\a_s$ and $\lambda$ will always refer to the running
coupling defined through the static force.
This scheme, also advocated in recent
years by Sommer \cite{Sommer}, has the obvious advantage that
the running coupling so defined doesn't have to change sign.  
The corresponding Gell-Mann-Low function, defined as
\begin{eqnarray}
\psi(\lambda(R))\equiv -R{d\lambda\over dR},
\end{eqnarray}
is given to three loops for $N_f=0$ but for any $N_c$ by \cite{NS}
\begin{eqnarray}
\psi(\lambda)\equiv -{11\over6}\lambda^2-{17\over12}\lambda^3
-(3.795...)\lambda^4-\cdots
\end{eqnarray}
The exact coefficient in the last term is rather cumbersome, so we
have merely quoted its numerical value to 3 decimal places.

\FIGURE[h!]{
\centerline{\scalebox{0.55}{\rotatebox{270}{\includegraphics{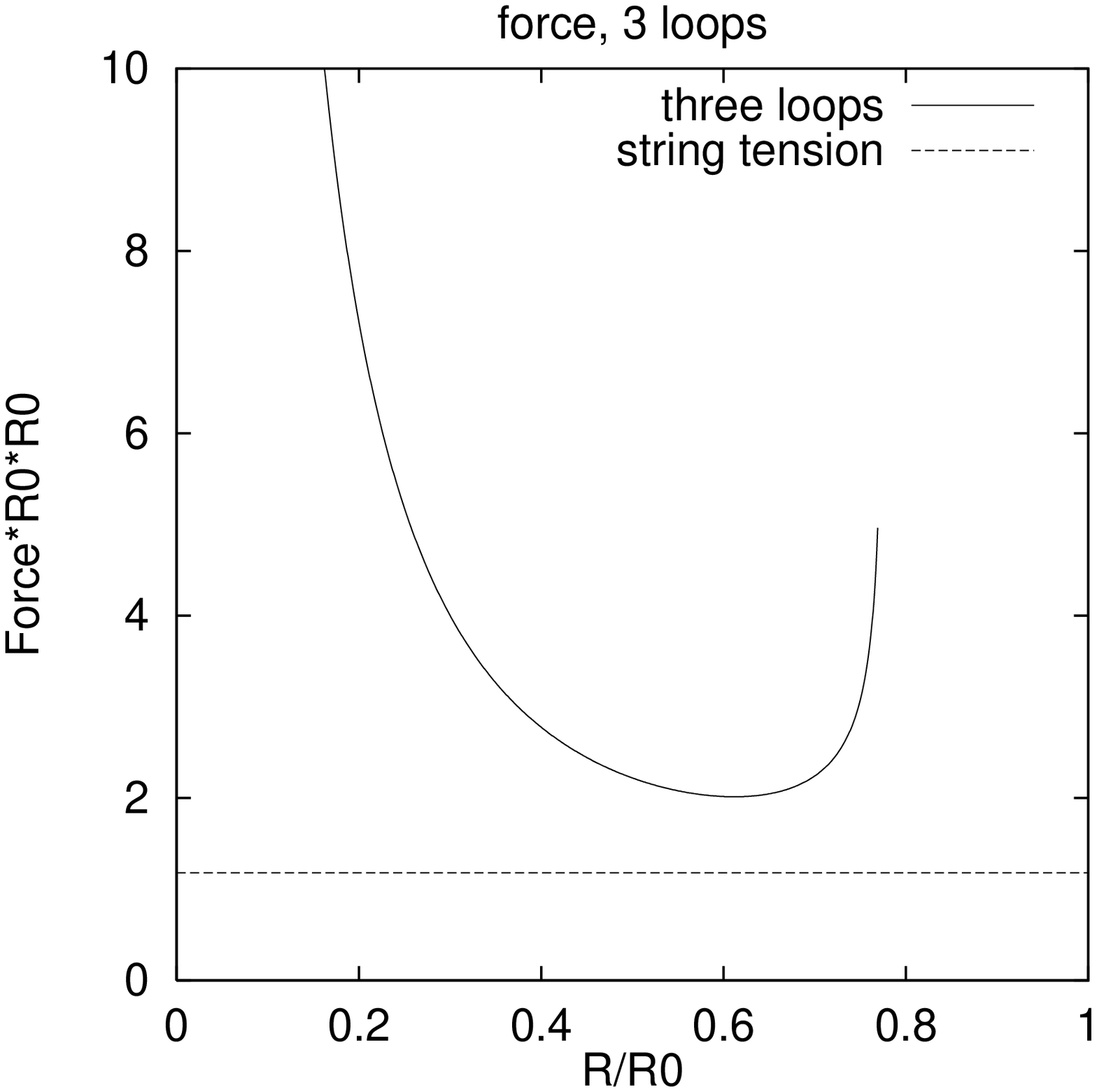}}}}
\caption{The force between static quarks in units of the Sommer scale
$R_0 \approx 0.5$ fm, computed perturbatively
to three loops at $N_c=3$ in the force renormalization
scheme \cite{NS}.  Also shown is
the asymptotic string tension of $(430 \mbox{MeV})^2 \times R_0^2$.}
\label{force}}

   Calculating $\lambda(R)$ by truncating $\psi$ at three loops leads
to a force that, as $R$ increases, drops to a minimum at $R_m$ given by
\begin{eqnarray}
   2 \lambda(R_m)+\psi(\lambda(R_m))=0,
\label{rmin}
\end{eqnarray}
after which the force {\it increases}, blowing
up as $(R_\infty-R)^{-1/3}$ at the Landau singularity $R=R_\infty$, 
as shown (for $N_c=3)$ in Fig.\ \ref{force}.
In this figure, $R_0$ is the Sommer scale $R_0 \approx 0.5$ fm. 
The corresponding static quark potential, in the force renormalization
scheme, is obtained by integrating the force,
\beq
    V_{q\bar q}(R) = V_{q\bar q}(R_A) + \int_{R_A}^R dR ~ \Bigl| F(R) \Bigr|
\label{integrate}
\eeq
where $R_A \ll R_0$ and $V_{q\bar q}(R_A)$ may be estimated
at sufficiently small $\a_s(R_A)$ by using eq.\ \rf{Vscheme},
with $\a_s^V\approx \a_s$.
Note that the Landau singularity in the force is integrable as $R\to
R_\infty$ from below, and the integrand becomes complex
for $R>R_\infty$. Thus the potential curve stops at $R_\infty$
at a finite value. 
It was found by Necco and Sommer in ref.\ \cite{NS} that the resulting
three-loop potential is a surprisingly accurate match to the lattice
Monte Carlo result, almost up to the Landau point at $R_\infty/R_0 \approx
0.78$.  This three-loop potential (with $R_A=0.15 R_0$) is displayed in Fig.\
\ref{potential}.
%\footnote{Note that truncation of $\psi$ at one or
%two loops leads to the same qualitative features of the force
%as seen in Fig.~\ref{force}. {\bf NEW FIGURE}.} 

%%%%

\FIGURE[h!]{
\centerline{\scalebox{0.55}{\rotatebox{270}{\includegraphics{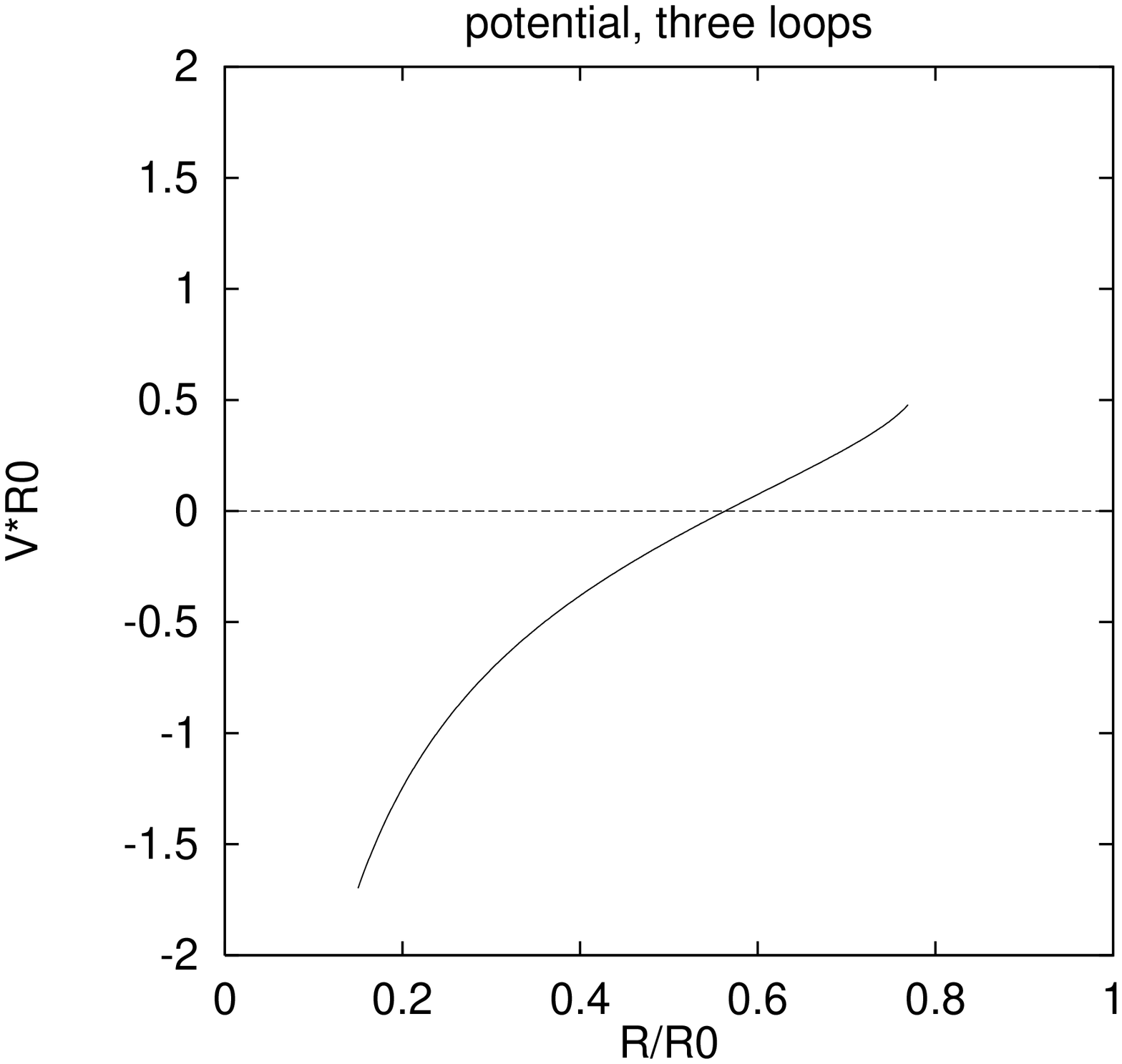}}}}
\caption{The static quark potential
at $N_c=3$, computed perturbatively to three loops 
in the force renormalization scheme.}
\label{potential}}

   It is clear 
%that the static potential increases faster
%than $-1/R$, which means 
that a potential of this sort can trap
a massless particle, whose kinetic energy $|{\bf p}|$ decreases as $1/R$
by the uncertainty principle: Minimizing $1/R +V(R)$ gives
\begin{eqnarray}
{1\over R^2}=V^\prime(R)=\left(1-{1\over N_c^2}\right){\pi\lambda(R)\over2R^2},
\end{eqnarray}
in other words $\lambda(R)=9/4\pi\approx0.71$ for $N_c=3$.
The effective coupling required for
binding a massless particle is of order $O(1)$, but not huge. 
There is thus hope that, after taking binding into account
with such an extrapolation of the RG improved
perturbative force, subsequent corrections may be under control.
It follows that, according to the force renormalization scheme, it is 
possible to have bound states containing massless constituent gluons,
such as glueballs and gluon chains.  Moreover, there is no indication
of any gross failure of perturbation theory up to $R\approx R_m$.  On the
contrary, as just mentioned, it seems that perturbation theory fits 
the numerical lattice data quite well. 

\subsection{Saturation Mechanism}

   Beyond $R=R_m$ the three loop force increases, and the
corresponding potential is concave upwards.  In contrast, a confining
force is expected to be asymptotically constant: One can say that the
perturbative force extrapolated to $R>R_m$ {\it over-confines}.  In
fact, there is a rigorous theorem, analogous to the well known
convexity of the thermal free energy, which states that the static
force never increases with quark separation, or, equivalently, that
the static potential is concave downwards \cite{Bachas}.  It appears
that physics beyond three loops, at $R>R_m$, must supply a mechanism
to \emph{weaken} the extrapolated perturbative force law, not
strengthen it.  In our opinion, this mechanism is supplied by the
formation of a gluon chain, as described in section 2.  As the
quark-antiquark separation increases, it can become energetically
favorable for a gluon to materialize between the quark and anti-quark.
Because the gluon is in the adjoint representation, its color can be
arranged so that it is simultaneously a sink of the color flux from
the quark and from the anti-quark, and indeed this is precisely the
arrangement of color dictated by the $N_c\to \infty$ limit.  The
effect of a gluon with this color orientation is to shield the direct
force of the quark on the antiquark, and thereby reduce the effective
separation of color charge. Further increase of $R$ leads to the
successive appearance of more constituent gluons, creating a gluon
chain between the quark and anti-quark.  Thus the separation of directly
interacting color charges never exceeds some maximum value.  It is
presumably this saturation mechanism which prevents the static force
from eventually increasing with quark-antiquark separation, in
accordance with the concavity theorem.  
The same saturation mechanism which bounds the
average color charge separation also bounds the corresponding effective
coupling.  If this maximal effective coupling is not too large, 
it should be possible to treat the interaction between neighboring 
gluons by perturbative methods.

In the next section we develop a variational approach to the
calculation of the properties of the gluon chain responsible
for quenching the increasing force beyond $R_m$. But it is
worth noting here that the form of the perturbative
force law shown in Fig.~\ref{force}, plus the hypothesis of a
saturation mechanism, already provides the basis for a
zero parameter estimate of the string tension, namely
$\s \approx F(R_m)$. In other words, we accept the extrapolated
perturbative force up to the point where it violates
the concavity theorem, after which we replace it 
with the simplest behavior consistent with the
theorem, namely a constant. This reasoning is reminiscent of the
Maxwell constuction in statistical physics, which restores
convexity to an approximate calculation of the free
energy of a Van der Waals gas. 

   As already noted, there is a theorem which tells us that the
increasing part of the force curve in Fig.\ \ref{force} (at $R>R_m$)
cannot apply to the actual force between static quarks.  However, if
that curve is taken as a paradigm of the nearest-neighbor force
binding gluons in the chain (neglecting change in gluon number), then
its increasing part may play a role in providing a restoring force
towards an equilibrium gluon separation, particularly if the mean
intergluon separation in the chain exceeds $R_m$.  In this case, there
is a net restoring force on any gluon fluctuating away from the mean
separation.  Again, the saturation mechanism will prevent any overall
increase of the static force with quark separation.

  To arrive at an estimate of the string tension, 
it remains to find $\lambda(R_m)$ 
and $R_m$ from Eq.~\ref{rmin}.
Truncating $\psi$ at three loops leads to a cubic equation
for $\lambda(R_m)$, with the numerical solution
$\lambda(R_m)\approx0.540$, so $(1-1/N_c^2)\pi\lambda(R_m)/2
\approx0.754$ for $N_c=3$. This value of $\lambda$ gives
the relative size of the terms entering into Eq.~\ref{rmin} as
\begin{eqnarray}
2\lambda-\psi_0\lambda^2-\psi_1\lambda^3-\psi_2\lambda^4
=.540(2-0.990-0.413-0.598),
\end{eqnarray}
so it is clear that one is extrapolating beyond the
strict validity of perturbation theory. 
However, if the three loop term turns out to be anomalously large,
the prospects for our perturbative approach would be brighter.
%Incidentally, with the three loop
%truncation of $\psi$, the force blows up as $C(R_\infty-R)^{-1/3}$
%at $R=R_\infty\approx1.266 R_m$.
From the slope of the meson Regge
trajectories $\alpha^\prime\approx0.86 {\rm GeV}^{-2}$ we have
the estimate from data of 
$\s=1/2\pi\alpha^\prime\approx
(430 {\rm MeV})^2$.  
This leads to $R_m\approx(495{\rm MeV})^{-1}\approx
0.40$fm. 
According to Figures 2 and 4 in ref.\ \cite{NS},
we can infer that with $\Lambda_{\overline{\rm MS}}=238 {\rm MeV}$,
the phenomenologically preferred value, $R_m\approx0.3$fm.
The string tension for this value would be higher by a
factor of $(1.33)^2\approx1.76$, 
\ie $\s \approx(570 {\rm MeV})^2$. 
Quark production, vetoed in our $N_c \ra \infty$ model, is expected to
reduce the predicted tension somewhat, but probably not this much.
Conversely, the string tension inferred from the meson
trajectories would require $\Lambda_{\overline{\rm MS}}=177$ MeV.

   Since the saturation mechanism must come into play at $R\approx R_m$, where 
perturbation theory begins to fail, we would conclude that the
QCD ground state $\Psi_{q\overline{q}}[A]$ in the presence of static
$q\overline{q}$ sources is dominated, at $R\approx R_m$, by a one-constituent
gluon component (perturbation theory expands around 
a state with zero constituent gluons).\footnote{
An old lattice Monte Carlo investigation of the gluon content of
the QCD string (second article of ref.\ \cite{Jeff}) suggests that
the one-constituent gluon component dominates beyond 3 lattice
spacings at $\beta=2.4$, corresponding to $R=0.73 R_0$.} 
If that is so, then the average separation of color charge along the
axis joining the static sources will be $R_m/2$.  Assuming this ``axial
separation'' of color charge remains fixed as quark separation 
$L$ increases, as it should according to the gluon chain model, then
for large separations, at $N_c=3$, we may estimate 
$L/N \approx R_m/2 \approx 0.15$ fm,  where $N$ is the number of 
gluons in the chain. This does not mean that the gluon is
sharply localized on the axis, however. Indeed, in order
to be bound it must be spread to a size $r$ at which the 
running coupling is of $O(1)$, which means $r > R_m$.

\section{A Variational Framework, and String-like Behavior}

   In order to make the notion of ``constituent gluon'' precise,
it is necessary to work in a Hamiltonian framework, which also means
picking a physical gauge, e.g.\ Coulomb or light-cone gauge.  The
ground-state wavefunctional in any physical gauge can be expressed in
the path-integral form
\beq
     \Psi_0[A(\vx)] = \int DA_\m(\vx,t<0) ~\d\Bigl[F(A)\Bigr] \D_{FP}[A]
        ~ \exp\Bigl[-\int_{-\infty}^0 dt L_A \Bigr]
\eeq
where $L_A$ is the gauge-field Lagrangian, $F(A)=0$ the gauge condition,
and $\D_{FP}$ the Faddeev-Popov determinant.\footnote{We note that in a
physical gauge either $\D_{FP}$ is trivial (as in light-cone gauge), so that 
there is no need to introduce ghost fields,
or else the ghost fields do not propagate in time, as in Coulomb gauge.}  
An excited state can
be constructed by multiplying the ground state with some polynomial in
the fields, i.e.
\beq
      \Psi_{ex}[A] = Q[A] \Psi_0[A] .
\label{excited}
\eeq
For example, the wavefunctional appropriate for a glueball state
would be 
\bea
      \Psi_G[A] = \sum_{N=1}^{N_{max}} \Psi^{(N)}_G[A]
\eea
where $\Psi^{(N)}_G[A]$ is the component of the glueball state with
$N$ constituent gluons, which has the general form
\bea
   \Psi^{(N)}_G[A] &=& \left\{ \int d\vx_1 d\vx_2 ... d\vx_N 
   ~f_{\m_1 \m_2 ... \m_N}(\vx_1,\vx_2,...,\vx_N)\right.\nonumber\\
   &&\hskip1cm  \left. \phantom{\int} \mbox{Tr}
    A_{\m_1}(\vx_1) A_{\m_2}(\vx_2) ... A_{\m_N}(\vx_N) \right\}
    \Psi_0[A] .
\eea
For our purposes, a ``constituent gluon'' simply refers to a gluon 
field operator multiplying the exact ground state.
The energy of any excited state \rf{excited} above the ground-state
energy, is given by
\beq
   E = {\langle \Psi_{ex}|H|\Psi_{ex}\rangle 
         \over \langle \Psi_{ex} |\Psi_{ex} \rangle } - \langle H \rangle .
\eeq
Defining
\beq
        Q_t \equiv Q[A(\vx,t)]
\eeq
the energy of the excited state can be computed from a correlation
function in the Euclidean-time version of the theory 
\beq
    E =  {-\oh} \lim_{T\ra 0} {d \over dT} \log
           \langle Q^\dg_T Q_{-T} \rangle 
\label{Eex}
\eeq
where $\langle Q^\dg_T Q_{-T} \rangle$ is the Euclidean vacuum expectation
value of these operators. In the perturbative calculation of $E$
one need only include the connected diagrams.

   Equation \rf{Eex} is the starting point for a field-theoretic
variational approach to bound states of constituent quanta \cite{GH,GH1}.  
The idea
is to first make some trial ansatz for $Q[A]$ involving a few
parameters, calculate the energy $E_{trial}$ of the trial state
perturbatively, via eq.\ \rf{Eex}, and then find the parameter values
which minimize this trial energy.  The result is a variational
estimate for bound-state energy, and an approximation to the
corresponding wavefunctional.

   In the case of the gluon chain, we take the static sources to be at
points $\vx = {\bf 0}$ and $\vx={\bf L}$, and consider $N$-constituent gluon
states generated from trial operators of the form
\bea
  Q_{chain}[A] &=& q^{a_1}({\bf 0}) \Bigl\{\int d\vx_1 d\vx_2 ... d\vx_N 
   ~\psi_{\m_1 \m_2 ... \m_N}(\vx_1,\vx_2,...,\vx_N) 
\non \\
    & &  A^{a_1 a_2}_{\m_1}(\vx_1) A^{a_2 a_3}_{\m_2}(\vx_2) ... 
        A^{a_N a_{N+1}}_{\m_N}(\vx_N) \Bigr\} \overline{q}^{a_{N+1}}({\bf L})
\eea
where $\psi$ is a trial N-gluon wavefunction.
%Tree-level contributions to the correlator are indicated schematically
%in Fig.\ xxx.  
In the $N_c \ra \infty$ limit, only interactions between
nearest-neighbor gluons (in the diagrams) need be taken into account.
A consistent chain picture will require that
the RG-improved intergluon interaction energy as
a function of gluon separation is qualitatively
similar to the static quark potential shown in
Fig.~\ref{potential}. Whether this happens remains to be seen.

\iffalse
Having obtained $\E(L)$ to some order in $\a_s$ , we then follow the
force renormalization prescription: Differentiate $\E(L)$ wrt $L$ to determine
the force and apply renormalization-group improvement.  The improved
force is then integrated to obtain an improved potential.
We then minimize the potential with respect to the variational parameters
to obtain the best estimate, in the variational approximation, to
the static quark potential.\\

\noindent{\bf Note}: This is probably over
simplified: RG improvement probably must be done internally
as well as externally.
\fi

\subsection{A simplified model}

   We will not attempt, in this article, a variational computation of
the gluon-chain energy $\E$ in the full field theory.  The perturbative 
evaluation of $\langle Q^\dg_T Q_{-T}\rangle$ for the gluon chain state
is very tedious, especially beyond tree level, and again
involves delicate issues of renormalization scheme dependence.
However, we can capture most of the important qualitative features of
the field theory calculation with a simple quantum-mechanical model,
in which the (N-1)-gluon Hamiltonian is taken to be
\beq
      H = \sum_{n=1}^{N-1} \Bigl|{\bf p}_n\Bigr| + 
          \sum_{n=2}^{N-1} V(\vx_n-\vx_{n-1}) + V_{qg}(\vx_1) 
          + V_{qg}({\bf L}-\vx_{N-1}) 
\eeq
where $V(x)$ is the gluon-gluon interaction energy,
while $V_{qg}(x)$ is the interaction energy between a 
static quark and its neighboring gluon. The restriction to 
nearest-neighbor interactions
is, of course, justified in the large $N_c$ limit.
In field theory the interaction between
each pair of neighbors is described by 
renormalization group improved Bethe-Salpeter style ladder
exchanges. In addition to the RG improved
Coulomb exchanges that build up the static force displayed
in Fig.~\ref{force}, exchange of transverse gluons
and contact interactions will add spin
and momentum dependence to the interaction. We can
anticipate that the most attractive spin channels are those
in which neighbor gluons are in a spin 0 state, so that for
a chain we should have anti-ferromagnetic order \cite{thorngluebag}. 
As we would do in the field theory calculation, 
we use the variational method in this
simplified model to obtain an approximation
to the N-gluon ground state energy ${\E}(L)$. 
%%%%%
As already mentioned, if the potentials $V$ and $V_{qg}$ have the qualitative
behavior shown in Fig.~\ref{potential}, a self-consistent gluon chain
will arise. \\ \\ \\ \\ \\ \\

\begin{itemize}
\item {\bf Product Ansatz $-$ Relative Coordinates}
\end{itemize}

   We begin with a simple product ansatz for the gluon-chain wavefunctional
\beq
\Psi(\vx_1,\vx_2,...,\vx_{N-1})=A\prod_{i=1}^N\psi({\bf u}_i)
\eeq
where the $\{ {\bf u}_i\}$ are relative coordinates
\beq
     {\bf u}_i = \vx_{i} - \vx_{i-1}
\eeq
and
\beq
     \vx_0 \equiv {\bf 0} ~~,~~~ \vx_N = {\bf L}
\eeq
It is convenient to also change integration variables to relative
coordinates, and the integration measure becomes
\beq
dV = {d^3 q\over(2\pi)^3}\prod_{i=1}^N d^3u_i \exp\left\{
i{\bf q}\cdot\left(\sum_{i=1}^N{\bf u}_i-{\bf L}\right)\right\},\qquad
\int dV |\Psi|^2=1
\eeq
where integration over $q$ gives a delta function enforcing
the constraint
\beq
\sum_{i=1}^N {\bf u}_i={\bf L}\equiv N{\bf R}
\eeq
Defining
\beq
F({\bf q})\equiv\int d^3u|\psi({\bf u})|^2e^{i{\bf q}\cdot{\bf u}}
\eeq
and the Fourier transform of the relative-coordinate wavefunction
\beq
\phi({\bf k})=
\int{d^3u\over(2\pi)^{3/2}}e^{-i{\bf k}\cdot{\bf u}}\psi({\bf u})
\eeq
we find for the normalization constant
\beq
A^{-2}=\int {d^3 q\over(2\pi)^3}e^{-i{\bf q}\cdot {\bf L}}F({\bf q})^N
\eeq
and the expectation value of the gluon nearest-neighbor potential
\beq
\langle{V({\bf u}_k)}\rangle =
A^2\int {d^3 q\over(2\pi)^3}e^{-i{\bf q}\cdot {\bf L}}
F({\bf q})^{N-1}\int d^3u|\psi({\bf u})|^2V({\bf u})
e^{i{\bf q}\cdot{\bf u}}
\eeq

   For the purpose of computing the expectation value of kinetic
energy, it is useful to introduce
\bea
\langle e^{i{\bf y}\cdot{\bf p}_k}\rangle&=&
A^2\int {d^3 q\over(2\pi)^3}e^{-i{\bf q}\cdot {\bf L}}F({\bf q})^{N-2}
\nonumber\\
&&\qquad\int d^3u d^3u^\prime e^{i{\bf q}\cdot({\bf u}+{\bf u}^\prime)}
\psi^*({\bf u})\psi^*({\bf u}^\prime)\psi({\bf u}+{\bf y})
\psi({\bf u}^\prime-{\bf y})\nonumber\\
&=&A^2\int {d^3 q\over(2\pi)^3}e^{-i{\bf q}\cdot {\bf L}}F({\bf q})^{N-2}
\nonumber\\
&&\qquad\int d^3k d^3k^\prime e^{i{\bf y}\cdot({\bf k}-{\bf k}^\prime)}
\phi^*({\bf k}+{\bf q})\phi^*({\bf k}^\prime+{\bf q})\phi({\bf k})
\phi({\bf k}^\prime)\\
\langle \delta({\bf p}_k-{\bf p})\rangle&=&
A^2\int {d^3 q\over(2\pi)^3}e^{-i{\bf q}\cdot {\bf L}}F({\bf q})^{N-2}
\nonumber\\
&&\qquad\int d^3k 
\phi^*({\bf k}+{\bf q}-{\bf p})\phi^*({\bf k}+{\bf q})\phi({\bf k}-{\bf p})
\phi({\bf k}) .
\eea
In the limit $N\to\infty$ we can evaluate the $q$ integral
by a saddle-point technique. Call ${\bf q}_0$ the saddle point
determined by
\begin{eqnarray}
{\nabla F\over F}({\bf q}_0)=i{{\bf L}\over N}=i{\bf R}.
\end{eqnarray}
Then we have in this limit
\begin{eqnarray}
\langle{V({\bf u}_k)}\rangle&=&
F({\bf q}_0)^{-1}\int d^3u|\psi({\bf u})|^2V({\bf u})
e^{i{\bf q}_0\cdot{\bf u}}\\
\langle e^{i{\bf Y}\cdot{\bf p}_k}\rangle&=&
F({\bf q}_0)^{-2}
\int d^3u\int d^3u^\prime e^{i{\bf q}_0\cdot({\bf u}+{\bf u}^\prime)} 
\nonumber \\
&&\qquad \psi^*({\bf u})\psi^*({\bf u}^\prime)\psi({\bf u}+{\bf Y})
\psi({\bf u}^\prime-{\bf Y})\\
\langle|{\bf p}_k|\rangle&=&
F({\bf q}_0)^{-2}\int d^3p|{\bf p}| \int {d^3Y \over(2\pi)^3} 
e^{-i{\bf Y}\cdot{\bf p}}
\int d^3u \int d^3u^\prime e^{i{\bf q}_0 
\cdot({\bf u}+{\bf u}^\prime)}
\nonumber\\
&& \qquad \psi^*({\bf u})\psi^*({\bf u}^\prime)\psi({\bf u}+{\bf Y})
\psi({\bf u}^\prime-{\bf Y})\nonumber\\
&=&F({\bf q}_0)^{-2}\int d^3p|{\bf p}|\int d^3k 
\nonumber\\ && 
\qquad \phi^*({\bf k}+{\bf q}_0-{\bf p})
       \phi^*({\bf k}+{\bf q}_0)\phi({\bf k}-{\bf p}) \phi({\bf k}) .
\end{eqnarray}

  Note that in our product ansatz for the trial wavefunction, all dependence 
on the gluon number in the chain, indicated by the
subscript $k$, has disappeared in $V({\bf u}_k)$ and $|{\bf p}_k|$.
The total energy of the trial state is then simply \footnote{We disregard
the distinction between $V$ and $V_{qg}$ at the ends of the chain, on 
the grounds that this is unimportant for ${\cal E}/L$ at large $L$.}
\beq
{\cal E}=N(\langle|{\bf p}_k|\rangle+\langle{V({\bf u}_k)}\rangle).
\eeq
Once this quantity is minimized as a function of the variational
parameters in the wavefunction, our estimate for the QCD string tension
is the energy per unit length
\beq
\s={{\cal E}\over L}={\langle|{\bf p}_k|\rangle+\langle{V({\bf u}_k)}
\rangle\over R}.
\eeq
To go further, we need to choose a definite $\psi({\bf u})$, for
example the gaussian 
\beq
\psi({\bf u})=e^{-u^2/2r^2} .
\eeq
In this case there are two variational parameters.  One of them
is the parameter $r$ in the above gaussian.  The other is the
number $N$ of gluons in a chain between heavy sources separated by distance
$L$ or, equivalently, the distance $R=L/N$.  For the gaussian wavepacket
we find
\bea
\phi({\bf k})&=&r^3e^{-k^2r^2/2}
\non \\
F({\bf q})&=&(\pi r^2)^{3/2}e^{-q^2r^2/4},
\qquad {\bf q}_0=-{2i{\bf R}\over r^2}
\non \\
\langle V\rangle&=&{1\over Rr\sqrt\pi}\int_0^\infty udu V(u)\left(
e^{-(u-R)^2/r^2}-e^{-(u+R)^2/r^2}\right)
\eea
for central $V({\bf u})$. In particular we find
\begin{eqnarray}
\left\langle {1\over |{\bf u}|}\right\rangle
={1\over Rr\sqrt\pi}\int_0^\infty du\left(
e^{-(u-R)^2/r^2}-e^{-(u+R)^2/r^2}\right)={\mbox{erf}(R/r)\over R}.
\end{eqnarray}
To evaluate the kinetic energy, note that
\begin{eqnarray}
\lefteqn{({\bf k}+{\bf q}_0-{\bf p})^2+({\bf k}+{\bf q}_0)^2+
({\bf k}-{\bf p})^2 +({\bf k})^2} 
\nonumber \\
  & & = 4{\bf k}^2+4{\bf k}\cdot({\bf q}_0-{\bf p})+
({\bf q}_0-{\bf p})^2+{\bf q}_0^2+{\bf p}^2\nonumber\\
& & = (2{\bf k}+{\bf q}_0-{\bf p})^2+{\bf q}_0^2+{\bf p}^2
\end{eqnarray}
so the ${\bf k}$ integral is a simple Gaussian leading to
\begin{eqnarray}
\langle|{\bf p}_k|\rangle&=&\left({r^2\over2\pi}\right)^{3/2}
\int d^3p|{\bf p}|e^{-r^2p^2/2}={1\over r}\sqrt{{8\over\pi}} .
\end{eqnarray}
Thus the product ansatz with gaussian wavepackets leads to
\beq
{{\cal E}\over L}={1\over rR}\sqrt{{8\over\pi}} + 
               {1\over R}\langle V \rangle .
\eeq
In particular, taking for $V$ the instantaneous Coulomb potential
\beq
      V(u) = -C_F {\a_s \over |u|}
\label{Coulomb}
\eeq
we have
\begin{eqnarray}
{{\cal E}\over L}={1\over rR}\sqrt{{8\over\pi}}
-{C_F\alpha_s \over R^2}
\mbox{erf}\left({R\over r}\right) .
\label{first_wf}
\end{eqnarray}

\begin{itemize}
\item {\bf String Wavefunction Ansatz}
\end{itemize}

   The second type of trial state we consider here is the
ground state wavefunction of a discretized string.  Strings with
discrete degrees of freedom were studied some time ago 
in ref.\ \cite{thornlcft}, and we will borrow
some results directly from that reference.

   The ground state of the discrete string is the state annihilated
by all the lowering operators for string modes
\beq
  a_m^i \Psi(\vx_1,\vx_2,...,\vx_{N-1}) = 0 ~~~~
    \left\{    \begin{array}{l}
         m=1,2,..,N-1 \cr
         i=1,2,3 \end{array} \right.
\eeq
where the position and momentum of the $k$-th gluon are related to
the string modes via
\begin{eqnarray}
{\bf x}_k&=&{{\bf L}\over N}k+\sqrt{2\over NT_0}\sum_{m=1}^{N-1}
{1\over\sqrt{2\omega_m}}\left({\bf a}_m+{\bf a}_m^\dagger\right)
\sin\left({m\pi\over N}k\right)\\
{\bf p}_k&=&-i\sqrt{2T_0\over N}\sum_{m=1}^{N-1}
\sqrt{\omega_m\over{2}}\left({\bf a}_m-{\bf a}_m^\dagger\right)
\sin\left({m\pi\over N}k\right)
\end{eqnarray}
where
$$\omega_m=2\sin{m\pi\over2N}$$
is the frequency of the $m$th mode. Then
we have
\begin{eqnarray}
\bra{0}e^{i{\bf x}\cdot{\bf p}_k}\ket{0}&=&\exp\left(-{\bf x}^2
\Delta^k_2\right)\nonumber\\
\bra{0}\delta({{\bf p}-{\bf p}_k})\ket{0}
&=&\left({1\over4\pi\Delta^k_2}\right)^{3/2}
\exp\left(-{{\bf p}^2\over 4\Delta^k_2}\right)\\
\bra{0}e^{i{\bf p}\cdot{\bf u}_k}\ket{0}&=&\exp\left(
i{\bf R}\cdot{\bf p}-{\bf p}^2\Delta^k_1\right)\nonumber\\
\bra{0}\delta({{\bf u}-{\bf u}_k})\ket{0}
&=&\left({1\over4\pi\Delta^k_1}\right)^{3/2}
\exp\left(-{({\bf u}-{\bf R})^2\over 4\Delta^k_1}\right)
\end{eqnarray}
where
\begin{eqnarray}
\Delta^k_1&=&{1\over NT_0}
\sum_{m=1}^{N-1}
{\sin{m\pi\over2N}}\cos^2\left({m\pi\over N}\left(k+{1\over2}\right)
\right)\nonumber\\
&=&{1\over4NT_0}\left\{\cot{\pi\over4N}%-1
+{1\over2}\cot{\pi(4k+3)\over4N}-{1\over2}\cot{\pi(4k+1)\over4N}\right\}\\
&\approx&{1\over\pi T_0}\left[1-{1\over(4k+3)(4k+1)}\right]
\to{1\over\pi T_0}\\
\Delta^k_2&=&{T_0\over N}
\sum_{m=1}^{N-1}
{\sin{m\pi\over2N}}\sin^2\left({m\pi\over N}k\right)\to{T_0\over\pi}\nonumber\\
&=&{T_0\over4N}\left\{\cot{\pi\over4N}%-1
-{1\over2}\cot{\pi(4k+1)\over4N}+{1\over2}\cot{\pi(4k-1)\over4N}\right\}\\
&\approx&{T_0\over\pi }\left[1+{1\over(4k+1)(4k-1)}\right]
\to{T_0\over\pi} .
\end{eqnarray}
From these expressions we find, for $V(u)$ proportional to $1/|u|$ as
in eq.\ \rf{Coulomb},
\beq
{{\cal E}\over L}={1\over R}{4\over\sqrt\pi}
\sqrt{\Delta_2}-{C_F\alpha_s \over R^2}
\mbox{erf}\left({R\over 2\sqrt\Delta_1}\right) .
\eeq

   Having tried two different types of trial wavefunctions,
the question is which leads to a lower ground state energy.
The answer is that the string wavefunction is the better of
the two, at least for inter-gluon potentials proportional to 
$1/|{\bf u}|$.  To see this, simply define $r\equiv 2\sqrt{\D_1}$,
so that
\beq
{{\cal E}\over L} =
{1\over rR}{8\over\sqrt\pi}
\sqrt{\Delta_1\Delta_2}-{C_F\alpha_s \over R^2}
\mbox{erf}\left({R\over r}\right)
\eeq
and, for the string wavefunction, $\Delta_1\Delta_2=1/\pi^2$.
Comparison with eq.\ \rf{first_wf} shows that for any values of
$r$ and $R$, the string wavefunction has a slightly lower energy
than the relative-coordinate product wavefunction.  

\subsection{String Tension}

   Without making any assumptions about the inter-gluon potential,
the energy per unit length of the gluon-chain in the 
string-wavefunction ansatz is
\beq
{{\cal E}\over L} = {1\over R}\left[ {8\over \pi^{3/2}}{1\over r} 
       + \langle V({\bf u}) \rangle \right] .
\label{EL}
\eeq
As explained in section 2, the minimal value of this energy per unit 
quark separation, as a function of the parameters $r,R$, is the variational 
estimate of the QCD string tension $\s$ in our quantum-mechanical model.
Of course, even the simplified model requires as input the intergluon
potential $V({\bf u})$, and to get this interaction energy right one
should probably use the variational approach in the full field theory.
However, even without knowing the intergluon potential precisely, we
can make some rough estimates using knowledge of the three-loop static
Coulomb potential as input.

  First of all, if $V(u)$ were simply due to tree-level effects, then
we would have
\beq
      \langle V({\bf u}) \rangle \propto \left\langle{1 \over |{\bf u}|}
                                      \right\rangle .
\eeq
Likewise, the running coupling is generally regarded as a function of the
\emph{inverse} color charge separation \cite{NS,Bali1}.
Let us therefore assume that $\langle V \rangle$ depends on the
variational parameters $r,R$ only through the VEV
\beq
      {1\over s} \equiv \left\langle{1 \over |{\bf u}|}
                                      \right\rangle
           = {1\over R} \mbox{erf}\left({R\over r}\right)
\eeq
so that we may write
\bea
         V(s) &=& \langle V({\bf u}) \rangle
\non \\
         F(s) &\equiv& {dV \over ds}
\eea
where $F(s)$ can be regarded as the magnitude of a semi-classical 
``force'' between gluons.

   Minimizing $\E /L$ results in two conditions.  The first, obtained
from minimizing wrt $R$, is
\beq
{1\over R}\left({8\over \pi^{3/2}}{1\over r} 
       + \langle V({\bf u}) \rangle \right) = {\pa \over  \pa R} V(s) 
\eeq
or, in view of \rf{EL},
\beq
      {\E \over L} =  {\pa \over \pa R} V(s) = {\pa s \over \pa R} F(s) .
\label{con1}
\eeq
This equation has a simple physical interpretation.  On the one hand,
the QCD string tension in the gluon chain model is simply the energy of
the chain per unit quark-antiquark separation, i.e.\ $\E/L$ on the
lhs of \rf{con1}.  On the other hand, the ``tension'' in this system
should be related to the change in potential energy of the system 
with respect to small deformations $\d R$ in inter-gluon separation;
this is the ``restoring force'' in the system along the quark-antiquark
axis.  The minimum condition \rf{con1} equates these two types of expression
for the string tension.  The second condition for the minimum is obtained
from minimizing $\E/L$ wrt $r$, which gives
\bea
    {8\over \pi^{3/2}}{1\over r^2} =  {\pa s \over \pa r} F(s) .
\label{con2}
\eea
Solving \rf{con2} for $F(s)$ and inserting into \rf{con1} gives an
expression for the string tension ($\s=\E/L$)
\beq
     \s = {\pa s \over \pa R} \left({\pa s \over \pa r}\right)^{-1} 
                {8\over \pi^{3/2}}{1\over r^2} 
\label{sigma}
\eeq
where
\bea
    {\pa s \over \pa R}  &=&  \mbox{erf}^{-1}({R\over r}) -
     {2\over \sqrt\pi}{R\over r}\mbox{erf}^{-2}(R/r)\exp(-R^2/r^2)
\non \\
    {\pa s \over \pa r}  &=& {2\over \sqrt\pi}{R^2\over r^2}
         \mbox{erf}^{-2}(R/r)\exp(-R^2/r^2) .
\label{deriv}
\eea

  Given $F(s)$, we could use \rf{con1} and \rf{con2} to determine $R$ and $r$,
and then \rf{sigma} would give the string tension.  Of course, this procedure
assumes that eqs.\ \rf{con1} and \rf{con2} have a solution at finite $r, R$.  
For $F(s)$ determined at tree
level, without taking account of loop corrections or the running coupling
constant, eq.\ \rf{con2} at fixed $R$
will typically have a solution at $r=\infty$
(small $\a_s$), or $r\propto R$ (large $\a_s$). But then 
\rf{con1} will  only be solved at $R=\infty$ or $R=0$
respectively. Clearly, incorporating the running coupling is a crucial
ingredient to a self-consistent chain solution. 
Even though
the detailed strengths and shapes will be different, 
we can expect the RG improved forces
(with $\psi$ truncated at one, two or three loops),
corresponding to $V$ and $V_{qg}$ in attractive channels, 
to have the same qualitative
behavior as Fig.~\ref{force}: an initial decrease to a
nonzero minimum, and a blowup at finite $R$ due to the Landau singularity.
This qualitative behavior is sufficient to guarantee a
consistent chain solution. Quantitative accuracy relies on
the chance that the effective coupling required for the
solution is not too large.

%As we have noted a number of
%times, a running coupling is necessary in order that the falloff in kinetic
%energy, as the gluon separation increases, is balanced at some separation
%by the increase in the potential energy, resulting in a bound state.

   At tree-level, the inter-gluon force $F(s)$ has three components: the
Coulomb force, the magnetic exchange force, and the contact interaction.
Some previous variational \cite{GH} and  bag \cite{thorngluebag}  
calculations suggest that the magnetic exchange force between 
neighboring gluons with total spin 0 will be attractive, 
and roughly equal in magnitude
to the Coulomb force, while the contact interaction is repulsive and
(for glueball wavefunctions) about 60\% as large as the Coulomb force.
For an accurate estimate of all contributions, we must actually carry
out the calculation for the gluon-chain
by the field theory method outlined above.  However,
based on the work in refs.\ \cite{GH,thorngluebag}, we expect that
the sum of contributions results in a (renormalization-group improved) 
intergluon force of the form
\beq
          F(s) \approx \kappa C_F {\a_s(s) \over s^2}
\label{Fs}
\eeq
where $\kappa$ is a number which is probably between 
one and two.\footnote{At loop level,
there will also be corrections to the kinetic energy term, which tend
to increase the size of that term \cite{GH}.}
  
   Even without knowing $F(s)$ precisely,
we can still use the three-loop perturbative results for the static potential
to make some educated guesses about the size of
$R$ and $r$, and arrive at a ``ballpark estimate'' for $\s$.   
At the end of section 3, we argued for 
$R = R_m/2 \approx 0.3 R_0$ , on the grounds that a constituent gluon
should appear between the quark-antiquark pair when they are $L=R_m$ apart, 
since this is where
the perturbative expansion around a zero constituent gluon state seems to be
breaking down.  For the $r$ parameter, we just note that it is not reasonable
to have $s$ beyond the Landau singularity, because before that distance we would
expect more constituent gluons to have appeared and reduced 
the effective charge separation.  
On the other hand, the bound state condition will require a 
running coupling $\a_s(s)$ which is $O(1)$, otherwise
the gluon kinetic energy dominates, and there is no binding.  So $s$
cannot be much less than the Landau singularity.  A guess which is perhaps
not so far off is $s=R_m=0.6 R_0$ which, when combined with 
$R=R_m/2=0.3 R_0$, implies $r=0.63 R_0$.
These particular guesses happen to land close to the right
answer, i.e.
\beq
       \s = {1.26 \over R_0^2}
\eeq
where the accepted answer for the QCD string tension is 
$\s=(430 \mbox{~Mev})^2 = 1.18 / R_0^2$.  This rather close agreement 
is probably fortuitous,
since we are only making guesses for $r$ and $R$, but it does
show that our numbers are in the right ballpark, and there
is some hope that the eventual calculation of $\s$ will be
in reasonable agreement with 
the phenomenological value.  It is also worth noting that with 
these parameters,
the spread in the wavepacket of each constituent gluon
(set by $s$), is roughly twice the size of the average 
intergluon separation along the quark-antiquark axis which
is $R\approx 0.3 R_0$.  Thus the QCD string is rather thick even before 
roughening effects come into play.

   Continuing a little further, we note that the above
choice of parameters $R,r$ implies
\bea
      \a(s) &=& {1\over \kappa C_F} {4\over \pi} e^{R^2/r^2}
\non \\
            &=& {1.60 \over \kappa C_F}
\eea
If we use the 3-loop running coupling in ref.\ \cite{NS}, and $N_c=3$,
then $s \approx 0.6 R_0$ would require $\kappa \approx 2$, which is
probably a little large.  But none
of these numbers should be taken very seriously at this stage, and a
full field-theoretic calculation using the string wavefunction ansatz
is clearly required, before we can draw any quantitative conclusions.

\subsection{Roughening and the L\"uscher Term}

The phenomenon of roughening, \ie\ the logarithmic
growth with $L$ of the transverse size of the gluon chain, requires
the long range correlations contained in the string wave
function ansatz. For example, the product
ansatz would not display this effect.
Indeed it is easy to calculate
\begin{eqnarray}
\bra{0}{\bf x}_{k\perp}^2\ket{0}&=&{D-2\over2NT_0}\sum_{m=1}^{N-1}
{\sin^2(m\pi k/ N)\over\sin(m\pi/2N)}\\
&\sim&{D-2\over2\pi T_0}\ln N
={r^2(D-2)\over8}\ln{L\over R}.
\end{eqnarray}
These same correlations also produce
certain finite size effects, that are
sub-dominant for $L/R\to\infty$, in the trial energy.
These come from explicit $1/N$ corrections and from
the fact that $\Delta^k_{1,2}$ have mild $k$ dependence.

To estimate these effects we first note some sums:
\begin{eqnarray}
\sum_{k=1}^{N-1}\Delta_2^k&=&{T_0\over2}\sum_{m=1}^{N-1}\sin{m\pi\over2N}
\sim {T_0\over2}\left[{2N\over\pi}-{1\over2}-{\pi\over24N}+\cdots\right]\\
\sum_{k=1}^{N}\Delta_1^k&=&{1\over2T_0}\sum_{m=1}^{N-1}\sin{m\pi\over2N}
\sim{1\over2T_0}\left[{2N\over\pi}-{1\over2}-{\pi\over24N}+\cdots\right].
\end{eqnarray}
It is of interest to estimate the finite size contributions to the
energy with our string trial wave function. For this purpose the detailed
form of the kinetic and potential energy is largely irrelevant and
it is more efficient to  use an arbitrary kinetic $K({\bf p})$
and potential $V({\bf r})$ energy in our formulas. We then have,
after changing integration variables to dimensionless ones: 
\begin{eqnarray}
\langle K({\bf p}_k)\rangle = {1\over\pi^{3/2}}
\int d^3p e^{-{\bf p}^2}K\left(2{\bf p}
\sqrt{\Delta^k_2}\right)\\
\langle V({\bf u}_k)\rangle = {1\over\pi^{3/2}}\int d^3u e^{-{\bf u}^2}
V\left({\bf R}+2{\bf u}\sqrt{\Delta^k_1}\right).
\end{eqnarray}
Then calling the limiting values of $\Delta^k_{1,2}\to\Delta_{1,2}$,
we can expand the energy to first order in $\Delta^k_{1,2}-\Delta_{1,2}$,
and do the sums over $k$.
\begin{eqnarray}
{\cal E}&=&
\sum_{k=1}^{N-1}\langle K({\bf p}_k)\rangle 
+\sum_{k=1}^{N}\langle V({\bf u}_k)\rangle\nonumber\\
&=&(N-1){1\over\pi^{3/2}}\int d^3p\ e^{-{\bf p}^2}K\left(2{\bf p}
\sqrt{\Delta_2}\right)+N{1\over\pi^{3/2}}\int d^3u\ e^{-{\bf u}^2}
V\left({\bf R}+2{\bf u}\sqrt{\Delta_1}\right)\nonumber\\
&&+\sum_{k=1}^{N-1}{\Delta_2^k-\Delta_2\over\Delta_2}
{\sqrt{\Delta_2}\over\pi^{3/2}}\int d^3p\ e^{-{\bf p}^2}
{\bf p}\cdot\nabla K \nonumber\\
&&+\sum_{k=1}^{N}{\Delta_1^k-\Delta_1\over\Delta_1}
{\sqrt{\Delta_1}\over\pi^{3/2}}\int d^3u\ e^{-{\bf u}^2}
{\bf u}\cdot\nabla V\\
&\sim&(N-1){1\over\pi^{3/2}}\int d^3p\ e^{-{\bf p}^2}K\left(2{\bf p}
\sqrt{\Delta_2}\right)+N{1\over\pi^{3/2}}\int d^3u\ e^{-{\bf u}^2}
V\left({\bf R}+2{\bf u}\sqrt{\Delta_1}\right)\nonumber\\
&&+\left[1-{\pi\over4}-{\pi^2\over48N}\right]
{\sqrt{\Delta_2}\over\pi^{3/2}}\int d^3p\ e^{-{\bf p}^2}
{\bf p}\cdot\nabla K \nonumber\\
&&+\left[-{\pi\over4}-{\pi^2\over48N}\right]
{\sqrt{\Delta_1}\over\pi^{3/2}}\int d^3u\ e^{-{\bf u}^2}
{\bf u}\cdot\nabla V.\nonumber
\end{eqnarray}
In these expressions $\nabla$ is always the gradient with respect
to the {\it argument} of the function that follows it.
Next we minimize ${\cal E}/L$ at fixed $L$
with respect to the parameters $T_0$ and
$R=L/N$, and obtain  
\begin{eqnarray}
0&=&
{\sqrt{\Delta_2}\over\pi^{3/2}}\int d^3p\ e^{-{\bf p}^2}
{\bf p}\cdot\nabla K -
{\sqrt{\Delta_1}\over\pi^{3/2}}\int d^3u\ e^{-{\bf u}^2}
{\bf u}\cdot\nabla V\\
0&=&
-{1\over R^2}\left[\int d^3p\ e^{-{\bf p}^2}K\left(2{\bf p}
\sqrt{\Delta_2}\right)+\int d^3u\ e^{-{\bf u}^2}
V\left({\bf R}+2{\bf u}\sqrt{\Delta_1}\right)\right]\nonumber\\
&&\hskip2cm +{1\over R}\int d^3u\ e^{-{\bf u}^2}{{\bf R}\over R}\cdot\nabla
V\left({\bf R}+2{\bf u}\sqrt{\Delta_1}\right)
\end{eqnarray}
These two equations determine $T_0$ and $R$. When both are satisfied,
we can simplify our expression for the variational
energy:
\begin{eqnarray}
{\cal E}&=&
NR{1\over\pi^{3/2}}\int d^3u\ e^{-{\bf u}^2}{{\bf R}\over R}\cdot\nabla V
\left({\bf R}+2{\bf u}\sqrt{\Delta_1}\right)
-{1\over\pi^{3/2}}\int d^3p\ e^{-{\bf p}^2}K\left(2{\bf p}
\sqrt{\Delta_2}\right)\nonumber
\\
&&+\left[1-{\pi\over2}-{\pi^2\over24N}\right]
{\sqrt{\Delta_2}\over\pi^{3/2}}\int d^3p\ e^{-{\bf p}^2}
{\bf p}\cdot\nabla K(2{\bf p}\sqrt{\Delta_2})\\
&=&
{L\over\pi^{3/2}}\int d^3u\ e^{-{\bf u}^2}{{\bf R}\over R}\cdot\nabla V
\left({\bf R}+2{\bf u}\sqrt{\Delta_1}\right)
-{\pi R\over24L}
{\sqrt{\Delta_2}\over\sqrt{\pi}}\int d^3p\ e^{-{\bf p}^2}
{\bf p}\cdot\nabla K(2{\bf p}\sqrt{\Delta_2})
\nonumber\\
&&+\left[1-{\pi\over2}\right]{\sqrt{\Delta_2}\over\pi^{3/2}}
\int d^3p\ e^{-{\bf p}^2}
{\bf p}\cdot\nabla K-\int {d^3p\over\pi^{3/2}}\ 
e^{-{\bf p}^2}K\left(2{\bf p}
\sqrt{\Delta_2}\right).
\end{eqnarray}
The first two terms are the linear potential and L\"uscher terms respectively.
The last two terms are independent of $L$ and can be absorbed
in the source energy. For $K({\bf p})=|{\bf p}|$, as
appropriate for massless gluons, the integrals in the
last three terms can be immediately carried out, yielding
\begin{eqnarray}
{\cal E}&=&
{L\over\pi^{3/2}}\int d^3u\ e^{-{\bf u}^2}{{\bf R}\over R}\cdot\nabla V
\left({\bf R}+2{\bf u}\sqrt{\Delta_1}\right)
\non \\
   && \qquad -{\pi\over24L}{4R\over r\sqrt{\pi}}
-\left[1+{\pi\over2}\right]{4\over r\pi^{3/2}}.
\end{eqnarray}
For the bosonic string the factor multiplying $\pi/24L$
in the Luscher term would be $D-2=2$ in 4 dimensional
space time. With our estimate above of $R\approx r/2$ this
coefficient is estimated to be $2/\sqrt\pi\approx1.13$.
But these numbers are far too preliminary to arouse
disappointment in such a discrepancy, especially since
it is not even settled that the bosonic string result
is correct for QCD.

\section{Conclusions}
We have developed our picture of a gluon chain into
a viable calculational framework for the physics
of quark confinement which stays close to perturbative
ideas. The crucial observation is that the
RG improved force law shows behavior which can
be interpreted as over-confinement. This shows
that perturbative physics might contain the germ
of quark confinement which, in combination with
the gluon chain saturation mechanism, leads
to a concrete proposal for detailed calculations.

We sketched a field theoretic variational approach to the 
gluon chain wave function. For a preliminary estimate of the
the string tension and other features of
the QCD string, we studied a simplified model, which
replaced field theoretic exchange interactions with 
a quantum mechanical potential. This exercise shows
how the chain model incorporates the important 
effects of roughening and the Luscher term in
the quark potential. In addition, the numerical
estimate for the string tension was certainly in the
right ballpark.

The next stage of this project is to redo the
variational treatment completely in the
context of field theory. One issue left unresolved
is whether the RG improvement of the
interaction between neighbor gluons on the
chain shows the same qualitative behavior
as that of the RG improved static force. If
it does, the consistency of our physical
picture will be confirmed. 
Whether the detailed quantitative predictions
of our model will then turn out as accurate as we hope $-$ say to 15
or 20 per cent $-$ is not yet clear. 

  We should finally note that the gluon chain model, which is
formulated in terms of particle-like excitations, is not opposed 
to a description of quark confinement in terms of some special class 
of field configurations, e.g.\ center vortices or monopoles, which 
dominate the QCD vacuum.  Instead, these approaches should
be seen as complementing one another.  The center dependence of
the asymptotic string tension provides a good example.  In terms
of particle (gluon) excitations, we can easily understand that this dependence
is due to string-breaking by particle production, which results
in color screening of the higher-representation heavy sources 
by constituent gluons.  In the gluon chain model in particular, 
the process is illustrated in
Fig.\ \ref{double}.  On the other hand, if one tries to explain the
center dependence of large Wilson loops in terms of field fluctuations
affecting the Wilson loop holonomy, then it is clear that the area law
must be due exclusively to fluctuations in the loop holonomy
among the center elements of the gauge group.  
This leads (perhaps inevitably) to a picture of the QCD vacuum
as being dominated at large scales by center vortex configurations.
Thus we have both a particle (chain breaking) and field (center vortex)
explanation for the same phemonemon, namely, the N-ality dependence of
the asymptotic string tension.   These particle/field descriptions need not
contradict one another; they are more likely to be dual
descriptions of the same underlying physics.  

    The attractive feature of the gluon chain model is that it offers
a simple and concise account of so many features of the QCD confining
potential: linearity, Casimir scaling (at large $N_c$), center dependence,
roughening, and the L\"uscher term.  Beyond that, the model provides
a promising framework for quantitative calculation of the string tension and,
perhaps, low-lying masses.  Whether these calculations are practical,
and if so how the results compare with phenomenology, remains to be seen.

\acknowledgments{ Our research is supported in part by
the U.S.\ Department of Energy under grant DE-AC03-76SF00098,
and in addition by grants \ DE-FG03-92ER40711 (J.G.), and
\ DE-FG02-97ER41029 (C.B.T). C.B.T. also 
acknowledges the support of the Miller Institute for
Research in Basic Science.}
%\bibliography{../../larefs}
%\bibliographystyle{unsrt}
%\end{document}

\end{document}